\documentclass[final,1p,times]{elsarticle}
\usepackage[english]{babel}
\usepackage{blindtext}
\usepackage{amssymb,amsmath,latexsym}
\usepackage[varg]{pxfonts}
\usepackage{algorithm}
\usepackage{algorithmic}
\usepackage{graphicx}
\usepackage{epstopdf}

\usepackage{amssymb}
\journal{TeX.sx}

\begin{document}
\begin{frontmatter}

\title{New algorithm of cuff-tissue-artery system modeled as the space axisymmetric problem}

\author{Jiacheng Xu}
\ead{jamiechen\_ sjtu@sjtu.edu.cn}
\author{Dan Hu\corref{cor1}}
\ead{hudan80@sjtu.edu.cn}
\address{Institute of Natural Sciences, Shanghai Jiao-Tong University, 800 Dongchuan Road, Shanghai 200240, China}
\address{Department of Mathematics, Shanghai Jiao-Tong University, 800
Dongchuan Road, Shanghai 200240, China}
\cortext[cor1]{Corresponding author}

\begin{abstract}
In this paper, mathematical models for cuff-tissue-artery system
are developed and simplified into an axisymmetric problem in space. It is nonlinear properties of cuff and artery wall that make it difficult to solve elastic equations directly with the finite element method, hence
a new iteration algorithm derived from principle of virtual work is designed to deal with nonlinear boundary conditions. Numerical accuracy is highly significant in numerical simulation, so it is necessary to analyze the influence different finite elements and grid generation on numerical accuracy. By dimensional analysis,
it is estimated that numerical errors must be $O(10^{-5})cm$ or less. To reach desired accuracy, the number of grids using higher order elements becomes one-fourth as large as that using low order elements by convergence rate analysis. Moreover, dealing with
displacement problem under specific blood pressure needs much small grid size to make numerical errors sufficiently small, which is not taken seriously in previous papers. However, it only takes a quarter of grids or less for displacement change problem to guarantee numerical accuracy and reduce computing cost.
\end{abstract}


\end{frontmatter}

\section{Introduction}
Blood pressure is considered as a critical reference index for diagnosing and predicting cardiovascular diseases in the clinical field \cite{article.1}. The accurate measurement of blood pressure has always been one of the most heated topics in both industry and academia.
It is well-known that the current mainstream approach to measure blood pressure is non-invasive methods including auscultation method and oscillometric method, which both use a cuff in the measurement.

Auscultation method \cite{Allen,Hamidon,India} is detecting the appearance and disappearance of Korotkoff sound generated by dynamic deformation of the blood vessel when cuff pressure changes. Based on the change of Korotkoff sound, diastolic and systolic pressure can be determined. Nowadays, auscultation method is widely accepted as the golden standard due to high accuracy. However, Owing to strict technique and inconvenience to carry a mercury sphygmomanometer, this method cannot be commonly used in each family.

Oscillometric method, which was proposed firstly by Marey \cite{Marey}, can measure cuff pressure oscillations when cuff pressure changes from above systolic to below diastolic pressure. The current oscillometric algorithms are mainly classified into amplitude-based method and slope-based method. Amplitude-based method usually uses maximum amplitude algorithm (MAA). It is now generally believed that this method can accurately estimate mean blood pressure, which can be explained by compliance of artery in some papers \cite{article.2,article.3,article.4}.
However, this method lacks of proper and precise methods to estimate systolic and diastolic pressure. Some researchers \cite{Geddes,Forster,article.5} obtain systolic and diastolic pressure by characteristic ratios while errors may be much large. The reason why the errors becomes large sometime is that characteristic ratios are acquired empirically even fixedly. Slope-based method is another way to get systolic and diastolic pressure by amplitude envelope slopes of oscillometric waveforms. This way can work greatly in clean data but fail in "dirty" data in reality because it is much sensitive to noise \cite{article.6}.

 Due to the limitations of traditional algorithms, alternative improved algorithms based on conventional oscillometric methods are presented, including Gaussian mixture regression approach \cite{article.7}, multiple linear regression and support vector regression approaches \cite{article.8}, neural networks \cite{article.9,For,article.10}, the least squares method \cite{article.11} and so on.

Previously, a large number of scholars have focused on oscillometric algorithms while theoretical analysis is extraordinarily deficient. To the best of the authors¡¯ knowledge, few investigators have begun to do theoretical research.
Jeon et al. used arterial-pressure-volume model to simulate oscillation waveform and proposed oscillometric algorithm using FFT \cite{article.17}.
Babbs incorporated artery wall model with cuff model to model cuff pressure oscillations and extracted systolic and diastolic pressure by the regression procedure \cite{article.12}.
These papers didn't take effects of tissue into consideration despite they simulated cuff pressure oscillations well. In fact, mechanical properties of tissue can have a significant influence on stress transmission efficiency.
Mauck et al. presented a simple one-dimensional cuff-arm-artery model to identify that the point of maximum oscillations can be used to determine mean blood pressure \cite{article.3}.
Ursino et al. developed a mathematical model of cuff-arm-artery system to study the influence of biomechanical factors on the measurement \cite{article.13}. Kim et al. combined vein model with cuff-arm-artery system to simulate the deformation of the arm tissue \cite{article.14}.
Liang et al. constructed three-dimensional cuff-arm-artery system incorporated with the bone to obtain stress distribution further explaining clinical observation \cite{Liang}.

In the present study, we develop a comprehensive cuff-tissue-artery coupled model and also simplify this model as a space axisymmetric problem. It is nonlinear properties of cuff and artery wall that make solving elastic equations more difficult. Inspired by principle of virtual work, we present a novel iterative algorithm to calculate elastic equations based on finite element method. Based on the problem we studied, we find numerical errors produced by mesh refinement should be $O(10^{-5}) cm$ or less by dimensional analysis. However, it is well-known that both mesh generation and element types
can affect numerical accuracy. Therefore, choosing proper finite element and grid size is so important in numerical simulations. It is found that compared with low order elements, the number of grids using higher order elements can reduce to one fourth since convergence rate is raised to about twice. What's more, it only takes a quarter of grids for displacement change problem compared with displacement problem under specific blood pressure.
That is, only small enough grid size can guarantee accuracy of solutions of displacements under specific blood pressure, which was always ignored in previous papers. However, dealing with displacement change just need fewer grids to meet precision requirements, which is beneficial to guarantee accuracy and save space in numerical research on cuff pressure oscillation later.
\section{Model building}\label{model}

As we know, it is geometry, mechanical properties and load conditions of the cuff-tissue-artery system that make it difficult to make a full analysis of the system. However, it is necessary to construct simple models under proper assumptions in order to understand the mechanism of the system.
In this problem, the effect of bone and body force are ignored. Although fat, muscle and other soft tissues have different mechanical properties such as elastic modulus and Poisson ratio, they can be seen as the same kind of isotropic and homogeneous elastic body with the same mechanical properties, that is, elastic modulus and Poisson ratio remain unchanged. Artery, tissue and cuff are axisymmetric with the same axis of symmetry, in which the artery is in the tissue and the tissue is wrapped with the cuff.

\subsection{Cuff modeling}\label{cuffmodel}
In the process of inflation and deflation of the cuff, the external wall of the cuff deforms much slightly, therefore, it is seen to be rigid. The internal wall of the cuff clings to the tissue tightly during the deformation of the cuff so that the compliance of it is seen to be large enough. Hence, cuff pressure is constantly the same as stress of the external tissue. The air inside the cuff is considered as ideal gas. Moreover, both inflation and deflation are seen as adiabatic processes satisfying
\begin{align}\label{cuffair}
(P_c+P_{atm})V_c^k=Q
\end{align}
where $P_c$ is cuff pressure, $P_{atm}$ is atmosphere pressure, $V_c$ is cuff volume, parameter $k=1.4$ describes adiabatic process of diatomic molecular gas, $Q$ denotes quantity of the remaining air inside the cuff at a specific time.

\subsection{Artery wall modeling}\label{arterymodel}
It is found experimentally that the cross-sectional area of artery will increase slowly with the increase of blood pressure until it reaches the maximal cross-sectional area in the state of expansion. That is, the compliance of artery decrease rapidly as the transmural pressure becomes larger when artery expands. In addition, the compliance of artery drops dramatically as the transmural pressure becomes a little smaller when the artery wall collapses. According to these experimental findings, some paper \cite{article.17} presented artery wall model described by compliance of artery, see equ. (\ref{artery}).
\begin{equation}\label{artery}
\left\{
\begin{aligned}
A_\alpha&=A_0\exp(C_aP_t),P_t<0\\
A_\alpha&=A_m-(A_m-A_0)\exp(C_bP_t),P_t\geq0. \\
\end{aligned}
\right.
\end{equation}
where $A_0$ is cross-sectional area of artery when transmural pressure becomes zeros, $A_m$ is the maximal cross-sectional area of artery, $C_a$ and $C_b$ are unknown parameters, describing the mechanical properties of artery. In simulations later, we use $C_a=0.09mmHg^{-1},b=0.03mmHg^{-1}$ which are the same as the paper \cite{article.17}.

\subsection{Tissue modeling}\label{tissuemodel}
In this problem, the geometry, external force distributions and constraints of tissue are symmetrical to an axis, therefore, this problem is seen as a space axisymmetric problem. In general, polar coordinates $r,\theta,z$ are much convenient than cartesian coordinates $x,y,z$ to solve this problem. Due to symmetry, the stress, strain and displacement at any point are independent of $\theta$, which are functions of $r$ and $z$. What's more, to keep symmetric, we can get that $\tau_{r\theta}=0,\tau_{z\theta}=0, u_\theta=0$.

In space axisymmetric problem, there are three kinds of equations to describe it, which are equilibrium equations (\ref{balance}), geometric equations (\ref{geo}) and physical equations (\ref{phy}), respectively.

Equilibrium equations:

\begin{equation}\label{balance}
\left\{
\begin{aligned}
\frac{\partial\sigma_r}{\partial r}+\frac{\partial\tau_{zr}}{\partial z}+\frac{\sigma_r-\sigma_\theta}{r}&=0\\
\frac{\partial\tau_{rz}}{\partial r}+\frac{\partial\sigma_z}{\partial z}+\frac{\tau_{zr}}{r}&=0. \\
\end{aligned}
\right.
\end{equation}

Geometrical equations:
\begin{align}\label{geo}
&\varepsilon_r=\frac{\partial u_r}{\partial r}~~~\varepsilon_\theta=\frac{u_r}{r}~~~\varepsilon_z=\frac{\partial w}{\partial z}~~~\gamma_{zr}=\frac{\partial w}{\partial r}+\frac{\partial u_r}{\partial z}
\end{align}

Physical equations:
\begin{equation}\label{phy}
\left\{
\begin{aligned}
\sigma_r&=\frac{E}{1+\mu}(\frac{\mu}{1-2\mu}\theta+\varepsilon_r)~~~\sigma_\theta=\frac{E}{1+\mu}(\frac{\mu}{1-2\mu}\theta+\varepsilon_\theta)\\
\sigma_z&=\frac{E}{1+\mu}(\frac{\mu}{1-2\mu}\theta+\varepsilon_z)~~~\tau_{zr}=\frac{E}{2(1+\mu)}\gamma_{zr}\\
\end{aligned}
\right.
\end{equation}

where $\theta=\varepsilon_r+\varepsilon_\theta+\varepsilon_z$, $E$ is elastic modulus and $\mu$ is Poisson ratio.
\section{Numerical algorithm}
The aim of this section is to solve coupled equations (\ref{balance}, \ref{geo}, \ref{phy}) numerically with the finite
element method. However, it is nonlinear boundary conditions that increases the difficulty of solving elastic equations. In order to deal with this problem, a proper iterative algorithm based on principle of virtual work is designed here.

From previous assumptions, body forces are ignored, hence an elastic body is just acted by surface forces.
For the elastic body with volume $\tau$ and
surface area $S$, $S$
is separated into force surface domain $S_\sigma$ and displacement surface domain $S_u$.
Surface forces, whose component is $X_i$, act on $S_\sigma$.
Displacements, whose component is $u_i$, are prescribed on $S_u$.
Assumed that there is any arbitrary virtual displacement in the body, the virtue work done by surface
forces on $S_u$ is zero since the
displacements are fixed on $S_u$. Without loss of generality, the area
in which surface forces work can be set to $S$. Therefore, the variation of work done by surface forces is
\begin{align}\label{work}
\delta W=&\int_{S}X_i\delta u_i\mathrm{d}S
\end{align}

Due to deformation, the variation of elastic potential energy of the body is
\begin{align}\label{strainpotential}
\delta U=&\int_{\tau}\sigma_{ij}\delta \varepsilon_{ij}\mathrm{d}\tau.
\end{align}

It is noted that stress tensor has a symmetrical property,
then substituting equ. (\ref{geo}) into equ. (\ref{strainpotential}), the variation of
elastic potential energy of the body can be rewritten as
\begin{align}\label{strainpotential1}
\int_{\tau}\sigma_{ij}\delta \varepsilon_{ij}\mathrm{d}\tau=&\int_{\tau}\frac{1}{2}\sigma_{ij}\delta(u_{i,j}+u_{j,i}) \mathrm{d}\tau=\int_{\tau}\frac{1}{2}\sigma_{ij}\Big((\delta u_i)_{,j}+(\delta u_j)_{,i}\Big) \mathrm{d}\tau\\\nonumber
=&\int_{\tau}\frac{1}{2}\sigma_{ij}(\delta u_i)_{,j}+\frac{1}{2}\sigma_{ji}(\delta u_j)_{,i} \mathrm{d}\tau=\int_{\tau}\sigma_{ij}(\delta u_i)_{,j}\mathrm{d}\tau.
\end{align}

Then, the variation of total energy is
\begin{align}\label{energy}
\delta E = &\delta U+\delta W\\\nonumber
 =&\int_{\tau}\sigma_{ij}(\delta u_i)_{,j}\mathrm{d}\tau+\int_{S}X_i\delta u_i\mathrm{d}S
\end{align}

Due to the equilibrium equations $(\sigma_{ij,j}=0)$, we have
\begin{align}\label{energy1}
(\sigma_{ij}(\delta u_i))_{,j}=&\sigma_{ij}(\delta u_i)_{,j}+\sigma_{ij,j}\delta u_i\\\nonumber
=&\sigma_{ij}(\delta u_i)_{,j}
\end{align}

Using Gauss' theorem, we have
\begin{align}\label{energy2}
\int_{\tau}(\sigma_{ij}(\delta u_i))_{,j}\mathrm{d}\tau=\int_{S}\sigma_{ij}\delta u_il_j\mathrm{d}S.
\end{align}
where $l_j$ is direction cosine of surface normal.

Therefore, we rewrite equ. (\ref{energy}) as
\begin{align}\label{energy3}
\delta E =&\int_{\tau}\sigma_{ij}(\delta u_i)_{,j}\mathrm{d}\tau+\int_{S}X_i\delta u_i\mathrm{d}S=\int_{S}\sigma_{ij}\delta u_il_j\mathrm{d}S+\int_{S}X_i\delta u_i\mathrm{d}S=\int_{S}\Big(\sigma_{ij}l_j+X_i\Big)\delta u_i\mathrm{d}S.
\end{align}

Total energy is the minimum when a elastic body is in the equilibrium state.
To make total energy reach minimum,
we choose $\delta u_i=-c\Big(X_i+\sigma_{ij}l_j\Big), c>0$ to
update $u_i$ so that $\delta E<0$.
Therefore, in each iteration we update displacements by $u_i=u_i-c\Big(X_i+\sigma_{ij}l_j\Big),c>0.$

\section{Results}
Before we make numerical experiments, we preset values of some parameters:
radius of external wall of cuff $R=2.7cm$, cuff pressure $P_c=100mmHg$, lengths of tissue and cuff are $L=20cm$ and $L_c=4cm$, elastic modulus of tissue $E=5*10^4Pa$, Poisson ratio of tissue $\mu=0.45$, systolic pressure $P_s=130mmHg$, diastolic pressure $P_d=70mmHg$, artery parameters $C_a=0.09mmHg^{-1},C_b=0.03mmHg^{-1}$, the radius of external tissue $r_{ext}=2.2cm$, the radius of artery under no force $r_0=0.2cm$, which is also the radius of internal tissue.

\subsection{Dimensional analysis}
Before numerical computation, we should do dimensional analysis to analyze numerical accuracy. Since the quantity of air inside the cuff remains almost unchanged in a much short time, we can deduce from cuff model $(P_c+P_{atm})V_c^k=Q$ that
\begin{align}\label{cuffapprox1}
(P_c+P_{atm})V_c^k&=(P_c+P_{atm}+\Delta P_c)(V_c-\Delta V_c)^k=(P_c+P_{atm}+\Delta P_c)V_c^k(1-\frac{\Delta V_c}{V_c})^k.
\end{align}

Due to the fact that $\frac{\Delta V_c}{V_c}<<1$, then neglecting the higher-order terms, equ. (\ref{cuffapprox1}) can be simplified into
\begin{align}\label{cuffapprox2}
(P_c+P_{atm})V_c^k&=(P_c+P_{atm}+\Delta P_c)V_c^k(1-k\frac{\Delta V_c}{V_c}).
\end{align}

Furthermore, equ. (\ref{cuffapprox2}) is simplified as follows when neglecting small quantity $\Delta V_c\Delta P_c$
\begin{equation}\label{cuffapprox3}
\Delta P_c=k(P_c+P_{atm})\frac{\Delta V_c}{V_c}.
\end{equation}

In a practical measurement, the amplitude of cuff pressure oscillation when cuff pressure is 100mmHg is about 1.5mmHg.
Under such circumstances, we can estimate that relative volume change rate is

\begin{equation}\label{RVCR}
\frac{\Delta V_c}{V_c}=\frac{\Delta P_c}{k(P_c+P_{atm})}=0.0012.
\end{equation}

To facilitate estimation, we assumed that the volume of the internal wall of the cuff always keeps cylindrical, hence $V_c\approx\pi R^2L_c-\pi r_{ext}^2L_c$. Moreover, Neglecting high-order term, it is estimated that $\Delta V_c\approx2\pi r_{ext}\Delta r_{ext}L_c$. Combining with equ. (\ref{RVCR}), we can get that displacement change of external tissue is
\begin{equation}\label{RVCR2}
\Delta r_{ext}=6.68*10^{-4}cm.
\end{equation}

From equ.(\ref{RVCR2}), we can know that the order of magnitude of
displacement change of external tissue is about $O(10^{-4}) cm$, therefore, in numerical computation, numerical errors produced by
mesh refinement should be smaller than it.

In fact, in using the finite element method, we need to choose proper elements and grid size to guarantee accuracy and reduce computational time and memory space. Here,
we compare two elements, which are low order elements and higher order elements. The main difference between them is higher order elements have
 quadratic displacement behavior since there is an intermediate node between the two nodes.

\subsection{Low order elements}\label{loworder}

\begin{figure}[h]
\begin{center}
\includegraphics[width=1.05\textwidth]{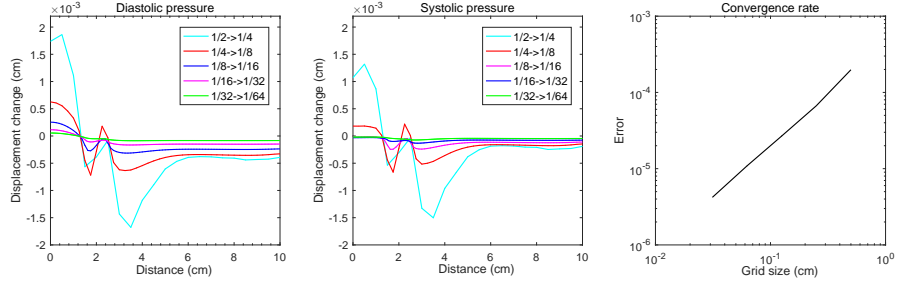} 
\caption{Left and middle: displacement change of external tissue by mesh refinement under diastolic and systolic pressure. (Symbol $\rightarrow$ represents grid size is refined) Right: convergence rate analysis of displacement change by mesh refinement under systolic pressure for low order element.}
\label{loworder_dispchange}
\end{center}
\end{figure}

From left and middle pictures in Fig \ref{loworder_dispchange}, we can find that under specific blood pressure,
the order of magnitude of displacement change of external tissue caused by mesh refinement  from the center of the cuff to the end is approximately $O(10^{-3}-10^{-4}) cm$ when grid size is larger than $1/32 cm$, but is
$O(10^{-5}) cm$ or less when grid size is less than or equal to $1/32 cm$.

Although the range of grid size meeting the predetermined accuracy is roughly known, there is still a question of how large grid size is needed to make numerical errors small enough.
To tackle this question, convergent rate is needed to be estimated so that proper grid size can be calculated.

Supposed that $u^{*}$ is true solution of elastic equations, $u_{2h}$ and $u_{h}$ are numerical solution when grid size are $2h$ and $h$, respectively, then we have
\begin{equation}\label{convergence}
\left\{
\begin{aligned}
u_{2h}-u^{*}&=C(2h)^\alpha\\
u_{h}-u^{*}&=Ch^\alpha\\
\end{aligned}
\right.
\end{equation}
where  $\alpha$ is convergent rate.

According to equ. (\ref{convergence}), we can know that
\begin{align}\label{convergence1}
u_{h}-u_{2h}&=h^\alpha(C-2^\alpha C).
\end{align}
where $\widetilde{C}=C-2^\alpha C.$

Taking norms on both sides of equ. (\ref{convergence1}), we can get
\begin{align}\label{convergence2}
\|u_{h}-u_{2h}\|&=\|\widetilde{C}\|h^\alpha.
\end{align}
Taking logarithms on both sides of equ. (\ref{convergence2}), we can get
\begin{equation}\label{convergence3}
log(\|u_{h}-u_{2h}\|)=\alpha log(h)+log(\|\widetilde{C}\|)
\end{equation}

From equ. (\ref{convergence3}), we can know that the relationship between $log(\|u_{h}-u_{2h}\|)$ and $log(h)$ is linear, and the slope is
convergent order $\alpha$. In addition, $\|\widetilde{C}\|$ can be calculated using the intercept.


From right picture in Fig \ref{loworder_dispchange}, we can calculate $\alpha=1.39$ and $\|\widetilde{C}\|=5.1971*10^{-4}$.
To reach the expected accuracy,
numerical errors produced by mesh refinement should be about $O(10^{-5})cm$, that is, $\|u_{h}-u_{2h}\|\sim O(10^{-5})cm$. Then we can estimate roughly that
$10^{-5}cm\sim \|\widetilde{C}\|h^{\alpha}$.
Therefore, $h\approx 0.0583cm$, that is, if numerical errors by mesh refinement is sufficiently small, grid size should
be less than or equal to $0.0583cm$, which is consistent with the results shown in the Fig \ref{loworder_dispchange}.

In fact, in our research, we pay more attention to cuff pressure oscillation or cuff volume oscillation during the change of blood pressure. Therefore, it is necessary to analyze numerical errors in numerical simulations to meet desired precision for cuff volume oscillation.
Here, we study displacement change of external tissue from diastolic to systolic, that is, displacement amplitude of external tissue in a cardiac cycle.

\begin{figure}[h]
\begin{center}
\includegraphics[width=1.05\textwidth]{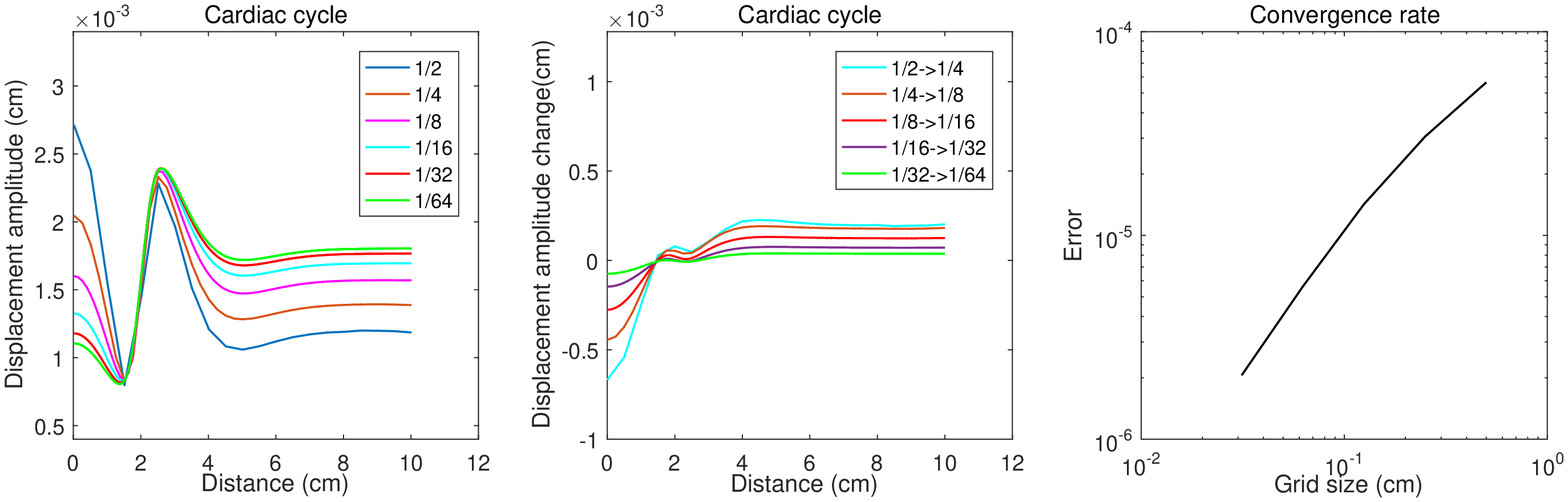} 
\caption{Left: displacement amplitude of external tissue under different grid size. Middle: displacement amplitude change of external tissue produced by mesh refinement. (Symbol $\rightarrow$ represents grid size is refined) Right: convergence rate analysis of displacement amplitude for low order element.}
\label{loworder_ampchange}
\end{center}
\end{figure}

From the left picture of Fig \ref{loworder_ampchange}, we can see that displacement amplitude of external tissue from the center of the cuff to the end in a cardiac cycle becomes closer when grid size gets smaller. In the middle picture, we can find that the order of magnitude of displacement amplitude change of external tissue caused by mesh refinement is
is approximately $O( 10^{-4}) cm$, which is a little better than that of displacement change produced by mesh refinement under specific blood pressure. Besides, numerical errors cannot decay to expected requirement until grid size is less than or equal to $1/16 cm$.

Then, we estimate how large grid size to achieve the precision by convergent rate. According to the right picture in Fig \ref{loworder_ampchange},
we calculate $\alpha=1.195$ and $\|\widetilde{C}\|=1.2948*10^{-4}$ for displacement amplitude problem.
Using similar method, we get step size $h\approx 0.1173cm$ by $10^{-5}cm\sim \|\widetilde{C}\|h^{\alpha}.$
Equally, if numerical errors by mesh refinement is sufficiently small, grid size should
be less than or equal to $0.1173cm$, which is also consistent with the results shown in the middle figure in Fig \ref{loworder_ampchange}.

\subsection{Higher order elements}
\begin{figure}[h]
\begin{center}
\includegraphics[width=1.05\textwidth]{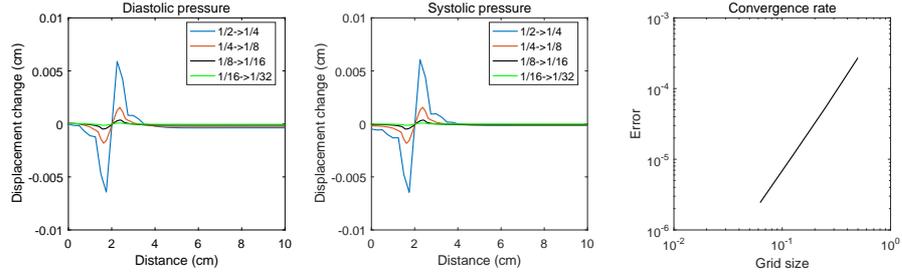} 
\caption{Left and middle: displacement change of external tissue by mesh refinement under diastolic and systolic pressure. Right: convergence rate analysis of displacement change for higher order element.}
\label{higherorder_dispchange}
\end{center}
\end{figure}

Now, we explore displacement change produced by mesh refinement under specific blood pressure using higher order elements. From left and middle pictures in Fig \ref{higherorder_dispchange}, it is found that the order of
magnitude of displacement change of external tissue from the center
of the cuff to the end caused by mesh refinement is approximately $O(10^{-3}-10^{-4}) cm$ when grid size
 is larger than $1/16cm$, but becomes
$O(10^{-5}) cm$ or less when grid size is less than or equal to $1/16cm$. Therefore, it will need about one quarter of grids when using higher order elements compared with using low order elements. That is, under specific blood pressure, convergence rate of higher order elements is about twice than that of low order elements.

Now, we can calculate convergent rate to verify the results and also estimate the proper grid size.
According to right picture in Fig \ref{higherorder_dispchange}, we can get convergence rate $\alpha=2.27$ and $\|\widetilde{C}\|=0.0013$,
subsequently, $h\approx 0.1172cm$ by $10^{-5}cm\sim \|\widetilde{C}\|h^{\alpha}$. Therefore, grid size is more than twice as big as that calculated with low order elements, which is consistent with the previous analysis.
Hence, as long as gird size is less than or equal to $0.1172cm$, numerical errors by mesh refinement for displacement problem under
specific blood pressure is about $O(10^{-5})cm$, which is similar as the results in the left and middle pictures in Fig \ref{higherorder_dispchange}.

\begin{figure}[h]
\begin{center}
\includegraphics[width=1.05\textwidth]{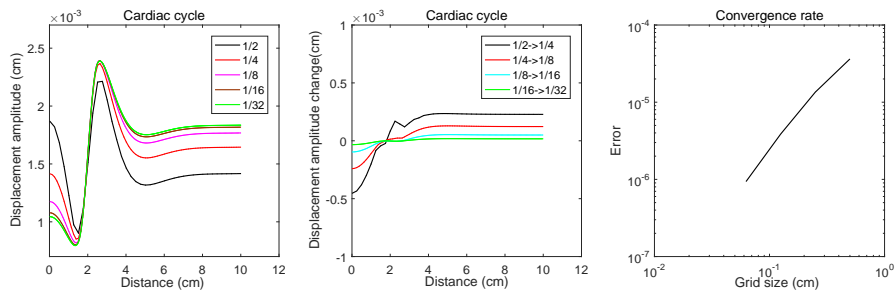} 
\caption{Left: displacement amplitude of external tissue under different mesh size. Middle: displacement amplitude change of external tissue produced by mesh refinement.Right: convergence rate analysis of displacement amplitude change for higher order element.}
\label{higherorder_ampchange}
\end{center}
\end{figure}

In the same way, we need to explore amplitude problem for higher order elements. As can be seen in Fig \ref{higherorder_ampchange}, displacement amplitude in a cardiac cycle becomes much closer when grid size gets a little smaller from left picture. In the middle picture, the order of
magnitude of displacement amplitude change by mesh refinement reaches $O(10^{-5}) cm$ fast as long as grid size is $1/8 cm$ or less. That is, the number of grids will be quartered compared with low order elements for amplitude problem.

Now, we also estimate convergence rate under such circumstances. From the right picture in Fig. \ref{higherorder_ampchange}, we can get $\alpha=1.76$ and $\|\widetilde{C}\|=1.2292*10^{-4}$. Subsequently,
we estimate grid size $h\approx 0.2404cm$ by $10^{-5}cm\sim \|\widetilde{C}\|h^{\alpha}$, which is also twice larger than that computed with low order elements.
In the same way, we see that if grid size is less than or equal to $0.2404cm$, numerical errors by mesh refinement for displacement amplitude will attain $O(10^{-5})cm$ or less.

\subsection{Remark}
No matter displacement problem under specific blood pressure or displacement change problem in a time period it is,
it will take a quarter of grids or less using higher order elements
to achieve desired accuracy compared with using low order elements. That is the advantage of higher order elements so that we will use them in numerical simulation later.

For displacement problem under specific blood pressure, mesh generation is greatly important because it needs a large number of grids to reach expected precision. In our analysis, to get desired accuracy, grid size tends to be much small even if higher order elements are used, i.e. $h\leqslant0.1172cm$. In fact, this problem is not taken seriously in previous papers.

However, for displacement change problem, there is an interesting result that we just use large grid size rather small grid size to meet the precision requirement. Specifically speaking, it is enough to reach the requirement when $h\leqslant0.2404cm$.

In fact, the focus of our research is cuff pressure oscillation of oscillometric method. Above all, the grid size of higher order elements is chosen to be $1/8cm$ for displacement change problem, so much less grids will be used and most importantly high accuracy can be attained in numerical simulation.

\section{Conclusion}
This paper presents space axisymmetric mathematical models of cuff-tissue-artery system based on proper assumptions and simplifications. Although elastic equations can be solved by the finite element method, nonlinear boundary conditions make it difficult to solve equations straightforwardly.

To deal with this tricky problem, this paper develops a new iterative algorithm inspired with principle of virtue work. Before we use the new iterative algorithm to do simulations, it is necessary for us to analyze numerical accuracy for the problem we studied. By dimensional analysis, the important finding is that numerical errors produced by mesh refinement for displacement change problem need to be much small, about $O(10^{-5})cm$ or less. Therefore, the chosen of finite element types and mesh generation is of greatly significance in numerical computation.

On the one hand, it is found that the number of grids using higher order elements can reduce to one fourth of that using low order elements no matter displacement problem under specific blood pressure or displacement change problem in a time frame it is. Hence, higher order elements will be chosed in numerical simulation later from this important result.

On the other hand, it will take very small grid size to meet precision requirements for displacement problem under specific blood pressure. In fact, few paper paid more attention to mesh generation when studying displacements under specific blood pressure, which may lead to large numerical errors using improper mesh generation. However, grid size can be more than doubled for displacement change problem. As a matter of fact, displacement change problem is exactly what we focus on. Therefore, large grid size will be used in the future work to reach desired accuracy and reduce computing cost.

\end{document}